\documentclass[fleqn,twoside]{article}
\usepackage{espcrc2}
\usepackage{graphicx}
\usepackage[figuresright]{rotating}

\newcommand{\AmS}{{\protect\the\textfont2
  A\kern-.1667em\lower.5ex\hbox{M}\kern-.125emS}}

\title{X-ray and radio prompt emission from a hypernova SN 2002ap }

\author{P. Chandra, \address{Joint Astronomy Programme, Indian 
Institute of Science, Bangalore;\\
 Tata Institute of Fundamental Research, Mumbai, \texttt{poonam@tifr.res.in}}
        A. Ray, \address{Tata Institute of Fundamental Research, Mumbai,
	\texttt{akr@tifr.res.in}}
	and
        F. Sutaria\address{Technical University Munich, Garching,
	\texttt{Firoza\_Sutaria@ph.tum.de}} \\
                }
       
\begin{document}

\begin{abstract}
Here we report on combined X-ray and radio observations of SN 2002ap with 
XMM-Newton ToO observation and GMRT observations aided with 
VLA published results. In deriving the X-ray flux of SN 2002ap we account for  
the contribution of a nearby source, found to be present in the
pre-SN explosion images obtained with Chandra observatory.
We also derive upper limits on mass loss rate from 
X-ray and radio data.
We suggest that the prompt X-ray emission is non-thermal in nature and 
its is due to the repeated compton boosting of optical photons.
We also compare SN's early radiospheric properties with two other 
SNe at the same epoch. 
\vspace{1pc}
\end{abstract}

\maketitle

\section{INTRODUCTION}
SN 2002ap was a type Ic SN discovered on Jan 29.4, 2002 in NGC 628. 
The explosion energy of the SN was larger than usual type Ic SNe 
(4-10$\,\times 10^{51}$ ergs).
SN 2002ap was of great interest because it was one of the closest 
extragalactic SN (D=7.3 Mpc); secondly it was a Type Ic SN and
 GRB association of type Ic 
is found in SN 1998bw (with GRB 980425) and SN 2003dh (with GRB 030329). 
SN 2002ap  has shown SN 1998bw like features like high velocity in early 
spectroscopy of optical data. 
 Study of type Ic SNe is interesting because they are devoid of H 
and He core and hence can probe closer to the  central engine.         

\section{OBSERVATIONS}
\subsection{X-ray Observations}
\begin{figure}
\includegraphics[width=4cm]{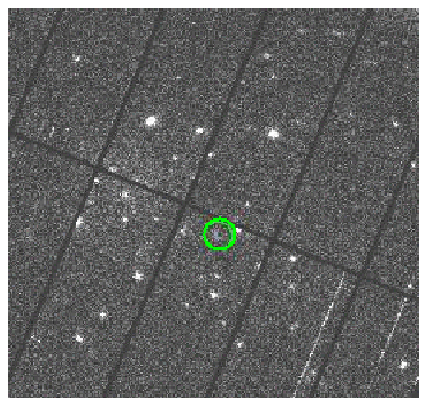}
\includegraphics[width=3cm]{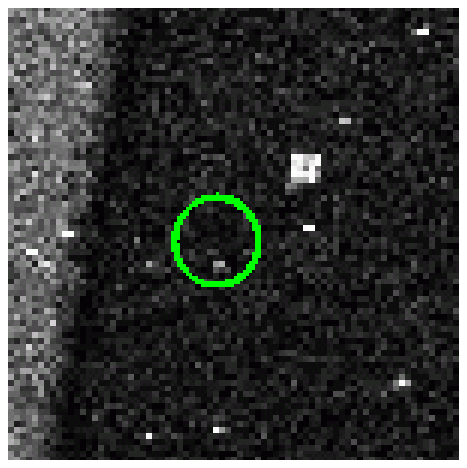}
\caption{LHS is XMM image of SN field. RHS is pre-SN Chandra field 
showing the presence of nearby object. Circle of $\sim$ 20"
is centered at SN.}
\label{fig: images}
\end{figure}

XMM-Newton observed SN 2002ap on Feb 2, 2002 for total 37.4 ksec with 
EPIC-PN and EPIC-MOS. We found
that Chandra Observatory has observed this field on Jun 19 and on 
Oct 19, 2001. Inspection of pre-explosion Chandra image reveals the
presence of a source 14.9" away 
from the SN. 
(See figure \ref{fig: images}).
A spectrum of the SN was extracted using circle 
of radius 40". While deriving the flux of the SN, we subtracted the 
contribution of nearby Chandra source. Table \ref{tab:5} \cite{sut03} 
gives results of 
different models fitted to the X-ray emission from the SN with a 
subtracted contribution due to Chandra source.
Both power-law and thermal bremsstrahlung models fit well. However
we cannot say which one is better due to sparse data.

\begin{table}
\tiny
\caption{Best fit spectral parameters for EPIC-PN data}
\label{tab:5}
\begin{tabular}{@{}llllll}
\hline
Model & $N_H$ & $\alpha$ & $kT$ & $\chi_{\nu}^2/dof$ & $f|_{0.3-10}$ \\
\hline
 & $10^{21}$ &  &  &  & $  10^{-14}$ erg \\
 & cm$^{-2}$ & & keV &  & /cm$^2$/s \\
\hline
Power-law & 0.49 & $2.6_{0.5}^{0.6}$ & --  & $ 1.2/20 $ & 1.07 \\
                                                                                
          & 0.42 & $2.5_{0.5}^{0.6}$ & --  & $ 1.2/20 $ & 1.0 \\
                                                                                
Thermal & 0.49 & -- & $0.84_{-0.3}^{+0.9}$ & $ 1.2/20$  &  0.81 \\
Brems. & & & & & \\
                                                                                
Raymond-  & 0.49 & -- & $2.31_{-0.8}^{+1.9}$ & $1.58/20$ & 1.04 \\
Smith     &       &    &                      &           &     \\
                                                                                
Blackbody & 0.49 & -- & $0.21_{-.06}^{+0.1}$ & $1.4/20$ & 0.6 \\
                                                                                
\hline
\end{tabular}
\end{table}

\subsection{Radio Observations}                                
The SN was observed with Giant Meterwave Radio telescope on Feb 5, 2002, at 
610 MHz and on Apr 08, 2002, in 1420 MHz band. The SN was quite weak in radio 
bands and we did not detect it at such low frequencies. Table 
\ref{tab:4} gives upper 
limit on the  fluxes of SN. The upper limits from GMRT at low frequencies
 combined 
with VLA detections at high freq. constrain models of early radio emission. 

\begin{table}
\caption{GMRT Observation log of SN 2002ap}
\label{tab:4}
\begin{tabular}{@{}lllll}
\hline
Date in       & $\nu$ & Resolution    & 2$\sigma$ Flux &  RMS  \\
2002     & (MHz)     & (arcsec)      & (mJy)  & mJy \\
\hline
                                                                                
5Feb    & 610       &9.5 x 6        & $< 0.34$         & 0.17 \\
8Apr    & 1420      & 8 x 3         & $< 0.18$         & 0.09 \\
\hline
\end{tabular}
\end{table}

\section{LOCATION OF THE RADIOSPHERE}

\begin{figure}
\includegraphics[width=7.5cm]{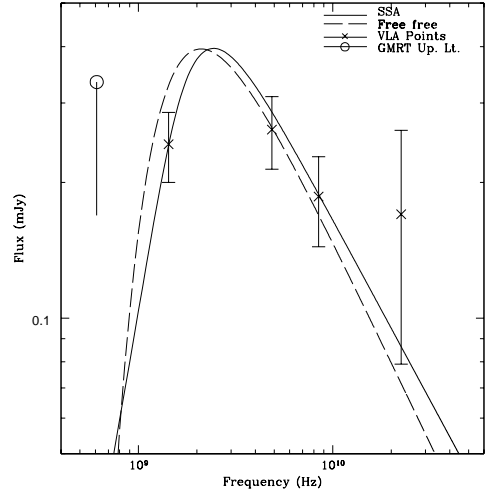}
\caption{Synchrotron self absorption and free-free fits to the
data on day 8.96 after the explosion. Magenta point is GMRT point whereas green ones are VLA points.}
\label{fig: 8day}
\end{figure}

We combined the high frequency VLA published data with low frequency GMRT 
data on the day 8.96 after the explosion and determined the location of the 
radio photosphere using the synchrotron self absorption fits. (Fig. 
\ref{fig: 8day}, Table \ref{tab:2}).

\begin{table}
\caption{Day 8.96 best fit SSA model to radio data}
\label{tab:2}
\begin{tabular}{@{}lllll}
\hline
$\alpha$   & $\nu_{p}$        & $F_{p}$       & $R_{r}$        & B    \\
         &GHz             & $\mu$Jy      & cm.          &  G    \\
                                                                                
\hline
0.8     & 2.45           & 397          & $3.5\times 10^{15}$ & 0.29 \\
\hline
\end{tabular}
\end{table}
XMM observed the SN in the UVW band with 
its on-board optical monitor system. The UVW1 flux for SN 
is of 7.667($\pm$ 0.002)$\times 10^{-15}$ erg cm$^{-2}$s$^{-1}$A$^{-1}$.
Table \ref{tab:maz}  reports the location of optical photosphere, which we 
calculated based on Mazzali et al's \cite{maz02} bolometric and
visual magnitudes on day 5.
We find that the radiospheric  velocity \cite{ber02} 
($\sim$90,000 km s$^{-1}$) is much higher 
than optical photosphere velocity (8000 km s$^{-1}$) on the same day.
This implies that the electrons responsible for the radio emission are
much farther away than the optical photosphere.

\begin{table}
\tiny
\small\caption{Optical obs. with XMM on board optical monitor}
\label{tab:maz}
\begin{tabular}{@{}llllll}
\hline
$M_{v}$   & $M_{bol}$        & $T_{eff}$       & $R_{opt}$   &$\bar{v}_{ph}$
 & $F_{UVOIR}$ \\
          &                 & K               & cm.             & km/s
    & $\rm erg cm^{-2} s^{-1}$\\
\hline
-17.4   & -16.5         & 11000         & $3.4\times 10^{14}$  & 8000       & $2\times 10^{-10}$\\
                                                                                
\hline
\end{tabular}
\end{table}

\subsection{Radio counterparts of X-ray sources}
\begin{figure}
\includegraphics[width=6.5cm]{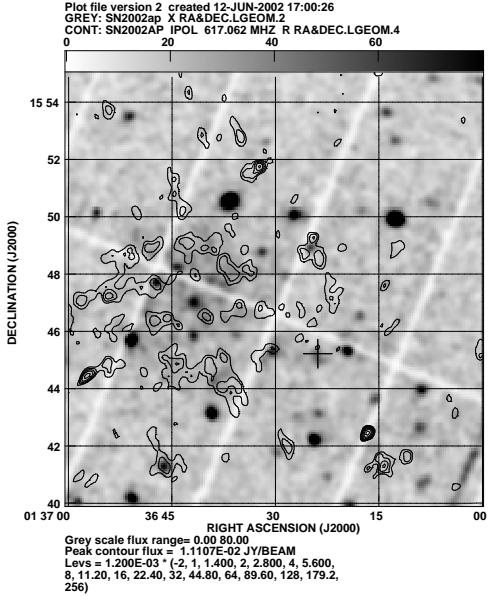}
\caption{The GMRT radio contours overlaid on XMM-Newton gray
scale image of SN 2002ap FOV. + shows position of SN.}
\label{fig: overlay}
\end{figure}

We overlaid the GMRT 610 MHz radio contours  on the XMM-Newton 
grey scale image of the SN 2002ap (see Fig \ref{fig: overlay}, Tab 
\ref{tab:3}). The image shows four X-ray sources 
having radio counterparts and table gives their co-ordinates.

\subsection{Mass Loss Rate from X-ray and radio}
We derive the upper limits on the mass-loss rate of the supernova
from both radio and X-ray datasets. 
For  photons of energy E$_{keV}$ , the time at which 
$\tau$=1 is given as
$$ t_X = C_{5} \dot M_{-5} u_{w1}^{-1} u_4^{-1} E_{keV}^{-8/3} $$
Since XMM observed X-rays from the SN on day 5 after, the explosion i.e. on 
t$_x$=5 the optical depth $\tau$ had reached 1. From this the upper limit
on the mass loss rate from X-ray observations is 
 $\dot M \leq \,4\,\times 10^{-5}\,M_{\odot}\,yr^{-1}$ for u$_{w}$=≈580
km s$^{-1}$ \cite{leo02}. 

Assuming that the dominant opacity source for radio radiation 
is free-free opacity
from fully ionised wind, the optical depth at radio frequency $\nu$ at time
t$_7=t/10^7$s is 
$$\tau_{ff}= 4 \dot M^2_{-5} u_{w1}^{-2} u_4^{-3} t_7^{-3} (\nu /1.4 GHz)^{-2}$$
From the radio detection at 1.4 GHz on day 5 the upper limit on the mass loss 
rate is  $\dot M \leq \,6\,\times 10^{-5}\,M_{\odot}\,yr^{-1}$.
Estimated from both radio and X-ray data are consistent with each other.

\begin{table}
\tiny
\caption{X-ray sources having radio counterparts}
\label{tab:3}
\begin{tabular}{@{}lllll}
\hline
Source  & RA            & Dec.          & Radio         & EPIC-PN \\
No.  &               &               &    Flux                & count-rate \\
        & J2000         & J2000         & (mJy)              & $10^{-3}$ ct s$^{-1}$ \\
\hline
1       & 01 36 47.2  & 15 47 45        & $7.1 (\pm 1.3)$    &  $4.3 (\pm 1.3)$
\\
2       & 01 36 46.1  & 15 41 17        & $12.8 (\pm 1.5)$   &  $3.9 (\pm 1.3)$
\\
3       & 01 36 24.9  & 15 48 58        & $22.7 (\pm 1.5)$   &  $1.9 (\pm 1.3)$
\\
4       & 01 36 30.5  & 15 45 17        & $4.5 (\pm 1.3)$    &  $4.6 (\pm 1.3)$
\\
\hline
\end{tabular}
\end{table}
\section{MECHANISMS TO PRODUCE X-RAY EMISSION}

In this section we evaluate  for different possible
mechanisms for the origin of the X-ray emission. 
We exclude the synchrotron origin of the X-rays because the direct
extrapolation of the radio flux up to X-ray frequencies, even
without a cooling break expected around optical band, leads to 
a flux of the order of few tens of picoJy, which is much smaller than 
the observed X-ray flux.

The observed X-ray flux could have been accounted by the thermal free-free
emission but that predicts  
a flat tail up to energy of $\sim$ 
100 keV which we don't see in our results. Hence we rule out the 
possibility of thermal free-free emission.

X-rays can be generated by repeated 
compton scattering by hot electrons off optical photons from the photosphere. 
The 2 fluxes (Compton and optical) are related by \cite{fra82},\cite{che94}
$$ {\cal F}_{\nu}^{Compton} \sim \tau_{e} {\cal F}_{\nu}^{opt}
(\nu_o/\nu)^{\gamma} \rm erg \; s^{-1} cm^{-2} Hz^{-1} $$
   where the optical depth is 

$$\tau_e  = {\dot M \sigma_T \over 4 \pi m_p R_s u_w} \Big(1 - {R_{opt} \over R_s}\Big)$$

and the energy index is

$$ \gamma (\gamma + 3) = -{m_ec^2\over kT_e} \rm ln\Bigl[{\tau_e\over
2}(0.9228- \rm ln \tau_e)
\Bigr] $$

The observed optical-UV flux and considerations of optical depth suggests that 
the above compton flux  can account for the observed X-rays; 
hence we suggest that the
early non-thermal X-ray flux is due to the compton scattering of the 
optical thermal photons.

Table \ref{tab:6} gives comptonizing plasma properties at t=5d for the 
two scenarios of the progenitor stars. The plasma has a 
maximum optical depth at 
twice the optical photosphere radius. The latter was taken as
$3.4\times 10^{14}$ cm. Since most of the X-ray emission would take place at
$\sim {\tau}_{max}$, the relevant plasma outflows with a velocity $ \sim$
16,000 km s$^{-1}$.
Both the progenitor scenarios can account for the observed optical
depth and hence can be a potential candidate for the progenitor 
system of the SN. 

\begin{table}
\tiny
\caption{Different progenitor Scenarios}
\label{tab:6}
\begin{tabular}{@{}lllll}
\hline
Scenario    & $\dot{M}_{-5}$  & $u_{w1}$ & $\tau_{e}$ &  $T_{e}$ \\
            & $10^{-5} \rm M_{\odot}/yr$ & 10 km/s &         &  $10^9$ K
           \\
\hline
Wolf-Rayet  & 1.5  & 58  & $4 \times 10^{-4}$ & $2$ \\
            & 3    & 100 & $4.4 \times 10^{-4}$ & $2$ \\
\hline
Interacting & 10 & 58 & $2.5\times 10^{-3}$ & $ 1.5$ \\
Binary      &    &    &                     &        \\
Case-BB$^a$      & 10 & 10 & $1.5 \times 10^{-2}$ & $1.1$ \\
\hline
\end{tabular}
$^a$\cite{hab}
\end{table}

\section{COMPARISON WITH OTHER SNE}

\begin{figure}
\includegraphics[width=7.5cm]{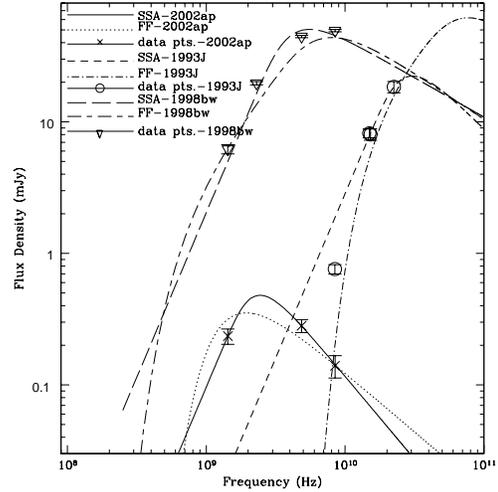}
\caption{Comparison of SN 2002ap with SN 1998bw and SN 1993J on day 11
after explosion. The solid line is SSA fit and dashed line is FF
fit.}
\label{fig: 3SN}
\end{figure}

We compare the spectrum of SN 2002ap with another type Ic SN 1998bw which 
had a GRB association and with a normal type IIb SN 1993J , on day 11 
after the explosion. Figure \ref{fig: 3SN} gives the comparison and
Table \ref{table:1} gives the best fit parameters. We observe that
SN 1998bw is moving with largest speed whereas the peak in the 
spectrum for SN 2002ap comes at lowest frequency.

\begin{table}
\tiny
\caption{The best fit table for the 3SNe}
\label{table:1}
\begin{tabular}{@{}lllllll}
\hline
SNe     & $\nu_{p}$ & $F_{p}$ &$\Theta_{eq}$ & 
$U_{eq}$ & $B_{0}$ & $R_{0}$ \\
\hline
       &          &         &               & $\times 10^{45}$
  &
& $\times 10^{15}$ \\
       &  GHz     & mJy     & $\mu$as      & erg                 & G       & cm
                 \\
\hline
2002ap & 2.45     & 0.48    & 39.0          & 0.69  & 0.47    & 4.80 \\
1998bw & 5.5      & 50.4    & 112.4         & 3500   & 0.23    & 68.4 \\
1993J  & 30.5     & 22.3    & 17.6          & 0.50  & 3.54    & 1.08 \\

\hline
\end{tabular}\\[2pt]
\end{table}

\section{DISCUSSION AND CONCLUSION}
 We suggest that repeated scattering of optical photons can account 
for the prompt 
X-ray emission from SN 2002ap. X-ray sphere (7$\times\,10^{14}$ cm) lies 
between optical (3.5$\times\,10^{14}$)and the radio photosphere 
(4$\times\,10^{15}$ cm). Both Wolf-Rayet and 
interacting binary case BB are capable of providing the adequate 
optical depth
 and are viable scenarios for the progenitor star. The properties of 
different kind of SNe differ significantly from each other. In fact the 
same type of SNe can show the significantly different properties.\\
\newline
{\bf Acknowledgment}
We thank staff of XMM-Newton and GMRT (NCRA-TIFR) that made these
observations possible.

\end{document}